\documentclass[11pt]{article}
\usepackage{amssymb}
\usepackage{graphicx}
\usepackage{lscape}
\usepackage{theorem}
[1999/03/29 PSNFSS v.7.2
Times font as default roman
: S Rahtz]

\newcommand \be{\begin{equation}}
\newcommand \ba{\begin{eqnarray}}
\newcommand \ee{\end{equation}}
\newcommand \ea{\end{eqnarray}}

\title{\LARGE Self-similar Approximants of the Permeability\\
in Heterogeneous Porous Media from Moment Equation Expansions}

\author{S. Gluzman$^{1}$ and D. Sornette$^{1,2,3}$\\
$^1$ Institute of Geophysics and Planetary Physics\\ University of
California Los Angeles, Los Angeles, CA 90095-1567\\
$^2$ Department of Earth and Space Sciences, UCLA\\
$^3$ Laboratoire de Physique de la Mati\`ere Condens\'ee\\ CNRS UMR
6622 and Universit\'e de Nice-Sophia Antipolis, 06108 Nice Cedex 2, France}

\begin{document}

\maketitle
\begin{abstract}
We use a mathematical technique, the self-similar functional
renormalization, to construct formulas for the average conductivity that
apply for large heterogeneity, based on perturbative expansions in powers of
a small parameter, usually the log-variance $\sigma_Y^2$ of the local conductivity. Using
perturbation expansions up to third order and fourth order in $\sigma_Y^2$
obtained from the moment equation approach, we construct the general
functional dependence of the transport variables in the regime where
$\sigma_Y^2$ is of order $1$ and larger than $1$. Comparison with available
numerical simulations give encouraging results and show that the proposed
method provides significant improvements over available expansions.
\end{abstract}

\pagebreak

\section{Introduction}

Subsurface hydraulic parameters such as medium permeability and porosity,
saturation curves and relative permeability have been traditionally viewed
as well-defined local quantities that can be assigned unique values at each
point in space. Yet, subsurface flow takes place in a complex environment
whose makeup varies in a manner that cannot be predicted deterministically
in all of its relevant details. This makeup tends to exhibit discrete and
continuous variations on a multiplicity of scales, causing hydraulic
parameters to do likewise. In practice, such parameters can at best be
measured at selected locations and depth intervals where their values depend
on the scale (support volume) and mode (instrumentation and procedure) of
measurement. Estimating the parameters at points where measurements are not
available entails a random error. Quite often, the support of measurement is
uncertain and the data are corrupted by experimental and interpretive errors.

Uncertainty is usually dealt with either deterministically through upscaling
or stochastically through the evaluating statistical moments. Statistical
moments can be obtained through Monte Carlo simulations or development of
moment differential equations, which is the method from which we start our
analysis.

In the stochastic approach, parameter values determined at various points
within a more-or-less distinct soil unit can be viewed as a sample from a
random field defined over a continuum. This random field is characterized by
a joint (multivariate) probability density function or, equivalently, its
joint ensemble moments. Thus, a parameter such as (saturated, natural) log
hydraulic conductivity $Y({\bf x})=\ln K_s({\bf x})$ varies not only across
the real space coordinates ${\bf x}$ within the unit, but also in
probability space (this variation may be represented by another
``coordinate'' $\xi$, the configuration coordinate, which, for simplicity,
we suppress). Whereas spatial moments are obtained by sampling $Y({\bf x})$
in real space (across ${\bf x}$), ensemble moments are defined in terms of
samples collected in probability space (across $\xi$).

In the moment equation approach, which we propose to exploit here, the
stochastic differential equations are averaged first to obtain moments
differential equations (MDEs) governing the statistical moments of the
dependent variables. The MDEs are themselves deterministic and can be solved
numerically or sometimes, analytically. The MDE approach has important
advantages. First, only a small number of equations must be solved: one for
the mean and one each for a small number of variances and covariances.
Second, the coefficients of the MDEs are relatively smooth because they are
averaged quantities. Thus the MDEs can be solved on comparatively smooth
grids. Third, the MDEs are available in analytical form, even though they
are usually solved numerically in applications. This holds the potential for
increased physical understanding of the mechanisms of uncertainty through
qualitative analysis. Finally, in many applications MDE approaches provide a
good estimate of the behavior of large variance systems despite being based
on small perturbation theory.

Since the moment equation approach derives the equations of evolutions of
the moments of the distribution of the transport variables by averaging the
stochastic differential equations of transport in heterogeneous porous
media, its fundamental limitation is the assumption that the variance $
\sigma_Y^2$ of the log hydraulic conductivity is small, in contradiction
with real geological settings of interest to a large variety of geophysical
applications. In order to obtain reliable descriptions of the transport
coefficients for large $\sigma_Y^2$, our goal in the present paper is to
adapt and extend the self-similar functional renormalization method to the
moment equation approach. In essence fundamentally non-perturbative, this
recently developed technique provides us with a stable and robust estimation of the
transport variables at large values of the perturbation parameter $
\sigma_Y^2 $. The functional renormalization method associates ideas from
the renormalization group theory of multiscale and critical phenomena \cite{mybook} with
methods from the theory of dynamical systems and of control theory. Using
perturbation expansions up to third order and fourth order in $\sigma_Y^2$
obtained from the moment equation approach, we construct the general
functional dependence of the transport variables in the regime where $
\sigma_Y^2$ is of order $1$ and larger than $1$.

The next section 2 recalls briefly how perturbation expansions are 
obtained from the moment equation approach. Section 3
summarizes the general formulation of the self-similar approximation theory.
Section 4 gives the results of the application of the functional renormalization
method to the moment equation expansions at increasing orders in $\sigma_Y^2$.
Section 5 formulates the expansion in powers of $1/d$, where $d$ is the space
dimension. This procedure well-known in statistical physics is resummed by
the fonctional renormalization to provide accurate formulas. Section
6 briefly outlines future directions of investigations.

\section{Problem Formulation and perturbation expansions}

\subsection{Basic equations}

It has become common to quantify uncertainty in ground water flow models by
treating hydraulic conductivity, $K$, and derived quantities like hydraulic
head, $h$, as random fields. For steady-state flows in the absence of
sources and sinks, the statistics of $h$ can be obtained from the stochastic
flow equation 
\begin{equation}
\label{fe}\nabla \,\cdot \,\left[ K({\bf x})\,\nabla h({\bf x})\right]
\;=\;0 
\end{equation}
when the statistics of $K$ are known. We further assume that the site of
interest is sufficiently characterized so that available experimental data
are sufficient to obtain the statistics of $K$, such as its (ensemble) mean, 
$\overline{K}$, variance, $\sigma _K^2$, and (two-point) correlation
structure, $\rho _K({\bf x},{\bf y})$. Then one can solve directly for the
moments of $h$ by developing deterministic equations for the moments from (
\ref{fe}). In general this involves taking the expected value of (\ref{fe})
and similar equations for higher-order moments, closing the system of moment
equations (usually through perturbation approximations). Numerical solutions
for moment equations are typically computationally more efficient than Monte
Carlo simulations. In the first place, taking expected values smoothes
parameters in the moment equations which in turn allows low-resolution grids
for numerical solutions. Furthermore, the number of moment equations is much
smaller than the number of realizations required by Monte Carlo simulations.
Additionally, the moment equations lend themselves to qualitative analysis.

Here, we concentrate on flow through highly heterogeneous porous media with
the variance $\sigma_Y^2$ of log hydraulic conductivity $Y = \ln K $ 
as the expansion parameter of the theory.
We estimate the mean hydraulic head, $\overline h({\bf x})$,
and assess the errors associated with such an estimation. We represent $K(
{\bf x}) = \overline{ K}({\bf x}) + K^{\prime}({\bf x})$ as the sum of a
mean, $\overline K({\bf x})$, and a zero-mean random deviation, $K^{\prime}(
{\bf x})$, with variance $\sigma_K^2({\bf x})$. Similarly, $h({\bf x}) = 
\overline{ h}({\bf x}) + h^{\prime}({\bf x})$ with $\overline{h^{\prime}}(
{\bf x}) \equiv 0$ and variance $\sigma_h^2({\bf x})$.

The average steady-state flow equation becomes 
\begin{equation}
\label{mfe}\nabla \,\cdot \,\left[ \overline{K}({\bf x})\,\nabla \,\overline{
h}({\bf x})\right] +\;\nabla \,\cdot \,\overline{{\bf r}}({\bf x})\;=\;0 
\end{equation}
which consists of a deterministic mean part, $\overline{K}\nabla \overline{h}
$, and a deterministic residual flux, $\overline{{\bf r}}=-\overline{%
K^{\prime }\nabla h^{\prime }}$. Solutions of (\ref{mfe}) require the mean
conductivity, $\overline{K}({\bf x})$, and in most cases, a method for
closing an expansion of $\overline{{\bf r}}({\bf x})$. Usually $\overline{
{\bf r}}({\bf x})$ is approximated through perturbation expansions based on $
\sigma _Y^2$, the variance of $Y=\ln K$, the logarithm of conductivity. This
approach works well as long as $\sigma _Y^2$ is small. This restriction is a
stumbling block on the road to applicability of numerous theoretical
analyses to real-world problems. Our goal here is to provide a general
theoretical method to extend the domain of application of moment equations
to the large heterogeneity limit.

\subsection{Perturbation Expansions}

Consider asymptotic expansions of the parameters and functions, $\overline K
\,=\, K_g\, (1 \,+\, \sigma_Y^2 / 2 \,+\, \ldots)$; $\overline {{\bf q}}
\,=\, \overline {{\bf q}}^{(0)} \,+\, \overline {{\bf q}}^{(1)} \,+\, \ldots$
; $\overline h \,=\, \overline h^{(0)} \,+\, \overline h^{(1)} \,+\, \ldots$
; and ${\bf r} \,=\, {\bf r}^{(1)} \,+\, \ldots$, where $K_g =
\exp(\overline Y)$, $\overline Y$ being the ensemble mean of $Y$. The
superscript $(i)$ denotes terms that are of $i$th-order, i.e. contain only
the $i$th power of $\sigma_Y^2$. The first-order (in $\sigma_Y^2$)
approximation of the residual flux is given by~\cite[and references therein]
{tart98a} 
\begin{equation}
\label{res}{\bf r}^{(1)}({\bf x}) \;=\; K_g\, \sigma_Y^2 \int_\Omega \rho_Y(
{\bf y},{\bf x})\, \nabla_{{\bf x}} \nabla_{{\bf y}}^T \,G({\bf y},{\bf x}
)\, \nabla \overline h^{(0)}({\bf y})\, d{\bf y}\, , 
\end{equation}
where $\rho_Y({\bf y},{\bf x})$ is the spatial two-point autocorrelation
function of $Y$, and $G({\bf y},{\bf x})$ is the deterministic Green's
function for Laplace equation in $\Omega$ subject to the corresponding
homogeneous boundary conditions. It is a standard practice in stochastic
hydrogeology to rely on the first-order approximation of ${\bf r}$~\cite
{dagan89}, but higher-order approximations are also available~\cite{hsu95}.

Collecting the terms of the same powers of $\sigma _Y^2$ yields the
zeroth-order approximation of the mean head in \ref{mfe}, 
\begin{equation}
K_g\,\nabla ^2\overline{h}^{(0)}({\bf x})\;=\;0\,, 
\end{equation}
and its first-order approximation, 
\begin{equation}
K_g\,\nabla ^2\overline{h}^{(1)}({\bf x})\;+\;\nabla \cdot \left[ \frac{
\sigma _Y^2}2\,K_g\,\nabla \overline{h}^{(0)}({\bf x})\;-\;{\bf r}^{(1)}(
{\bf x})\right] \;=\;0\,. 
\end{equation}
Solving a system of these sequential approximations leads to $\overline{h}
^{[1]}\,\equiv \,\overline{h}^{(0)}\,+\,\overline{h}^{(1)}$. Strictly
speaking, for such expansions to be asymptotic it is necessary that $\sigma
_Y^2\ll 1$, i.e. that porous media be mildly heterogeneous. However, various
numerical simulations (e.g.,~\cite{guad99b}) have demonstrated that these
first-order approximations remain remarkably robust even for strongly
heterogeneous media with $\sigma _Y^2$ as large as $4$.

\subsection{Effective Conductivity of Porous Media}

For an effective conductivity to exist in the strict sense, it is necessary
that $\nabla \overline h$ be constant. A somewhat a less restrictive
assumption requires $\nabla \overline h$ to vary slowly in space, i.e. to
have negligibly small derivatives~\cite{dagan89}. Then one can localize
expression (\ref{res}) as 
\begin{equation}
\label{resapprox}{\bf r}^{(1)}({\bf x}) \;\approx\; K_g\, \sigma_Y^2 {\bf A}
^{(1)}({\bf x})\, \nabla \overline h^{(0)}({\bf x})~, \hspace{1.5cm} {\bf A}
^{(1)}({\bf x}) \;=\; \int_\Omega \rho_Y({\bf y},{\bf x})\, \nabla_{{\bf x}}
\nabla_{{\bf y}}^T \,G({\bf y},{\bf x})\, d{\bf y} ~. 
\end{equation}

Under these conditions, retaining the two leading terms in the asymptotic
expansion of the mean Darcy flux, $\overline {{\bf q}} \,\approx\, \overline
{{\bf q}}^{[1]} \,\equiv\, \overline {{\bf q}}^{(0)} \,+\, {\bf q}^{(1)}$,
yields 
\begin{equation}
\label{eq}-\, \frac{\overline {{\bf q}}^{[1]}({\bf x})}{K_g} \;=\; \nabla
\overline h^{(1)}({\bf x}) \;+\; \left[ {\bf I} \,+\, \sigma_Y^2\, \left( 
\frac{1}{2}\, {\bf I} \,-\, {\bf A}^{(1)}({\bf x})\right) \right]\, \nabla
\overline h^{(0)}({\bf x})\,. 
\end{equation}
For flow through infinite, statistically homogeneous porous media under mean
uniform flow conditions, or at points away from boundaries and
singularities, the mean hydraulic head gradient $\overline {{\bf J}} \,=\,
\nabla \overline h^{(0)} \,=\, const$ and $\nabla \overline h^{(i)} \,=\, 0$
($i \,\ge\, 1$)~\cite{dagan89,tart98c}. This gives rise to the effective
conductivity given approximately by 
\begin{equation}
\label{k} K_{ef}^{[1]} \, \equiv\, K_{ef}^{(0)} \,+\, K_{ef}^{(1)} \,=\,
K_g\, \left[1 + \left(\frac{1}{2} - \frac{1}{d}\right)\, \sigma_Y^2\right] 
\end{equation}
where $d$ is the space dimension.

Various attempts to generalize this asymptotic expansion to highly
heterogeneous formations were attempted by conjecturing that expression (\ref
{k}) represents the two leading terms in the expansion of an exponent~\cite
{matheron67,shvidler62}, 
\begin{equation}
\label{kk}K_{ef}\;=\;K_g\,\exp \left[ \left( \frac 12-\frac 1d\right)
\,\sigma _Y^2\right] \,. 
\end{equation}
In recent years the question of validity of expression (\ref{kk}) was the
focus of a thorough investigation. It was proven that expression (\ref{kk})
is rigorously valid under one-dimensional flow in log-normal fields where it
yields the harmonic mean $K_h=K_g\exp (-\sigma _Y^2/2)$~\cite
{dagan93,paleologos96}. It is also rigorously valid under two-dimensional
flow in log-normal, statistically isotropic conductivity fields where it
yields the geometric mean $K_g$~\cite{matheron67}. For three-dimensional
flow in log-normal, statistically isotropic fields, the second-order (in $
\sigma _Y^2$) term in (\ref{k}) was found to be in agreement with the Taylor
series expansion of (\ref{kk})~\cite{dagan93}. While unsuccessfull attempts
to prove (\ref{kk}) for three-dimensional flows in such fields have been
reported~\cite{king89,noetinger90}, De Wit~\cite{dewit95} demonstrated that
the third-order correction in (\ref{k}) is not equal to the third-order term
in the Taylor expansion of ( \ref{kk}), thereby proving this conjecture to
be not strictly valid for three-dimensional Gaussian isotropic media.
Instead, it was demonstrated that this and higher-order terms depend on the
shape of the correlation function $\rho _Y$.

These results suggest that it would be beneficial to view the equation (\ref
{k}) and its higher order terms as a perturbation expansion of the true
transport variable in powers of the variance $\sigma _Y^2$. In this sense,
the passage from (\ref{k}) to (\ref{kk}) is a resummation procedure. It thus
makes full sense to ask what could be the most general and robust
resummation that can generalize (\ref{k}) in order to extend its domain of
validity in the regime of large $\sigma _Y^2$ where the initial perturbation
expansion breaks down.

It is often the case that perturbation expansions are not converging but are
instead diverging series. Even if the series is convergent for small
perturbation parameters $\sigma _Y^2$, one is in general interested in the
regime where $\sigma _Y^2$ is of order $1$ and larger. In this case, the
perturbation series is divergent and is of no direct use. The study of such
summation of divergent series is the problem of great importance in
theoretical physics, applied mathematics and engineering. This is because
realistic problems are usually solved by means of some calculational
algorithm often resulting in divergent sequence of approximations. Assigning
a finite value to the limit of a divergent sequence is called
renormalization or summation technique. The most widely used such technique
is Pad\'e summation \cite{B}. However, the Pad\'e
summation method has several shortcomings. First of all, to reach a
reasonable accuracy of Pad\'e approximants, one needs to possess tens of
terms of a perturbation series. In contrast, only a few terms are often
available because of the complexity of the problem. Second, Pad\'e
approximants are defined for the series of integer powers. But in many
cases asymptotic series arise having noninteger powers. Third, there are
quite simple examples that are not Pad\'e summable even for a
sufficiently small variable. Last but not least, Pad\'e
summation is more of a numerical technique providing answer in the form of
numbers. Therefore, it is difficult, if possible, to analyze the results
when the considered problem contains several parameters to be varied, since
for each given set of parameters one has to repeat the whole procedure of
constructing a table of Pad\'e approximants and of selecting
from them one corresponding to a visible saturation of numerical values.

We thus turn to the method of so-called self-similar approximation or
functional renormalization that provide a very interesting alternative. We
first summarize the idea of the technique and then apply it to calculate
properties of transport in porous media in the limit of large heterogeneity.

\section{General formulation of the self-similar approximation theory \label
{inrol}}

General ideas and the mathematical foundation of the self-similar
approximation theory have been described in detail in \cite
{G1,G2,G3,G4,G5,G6,G7,G8,G9,G10,G11}. The approach is applicable in all
cases, when either just a few terms of a series are known or when a number
of such terms are available. We are always able to obtain analytical
formulas that are easy to consider with respect to varying characteristic
parameters. We now expose the general idea of the method of self-similar
approximation.

Consider the case, when for a sought function $f(x)$, one derives an
approximate perturbative expansion 
\begin{equation}
\label{f1}p_k(x)=\sum_{n=0}^ka_n\ x^{\alpha _n}, 
\end{equation}
in which $\alpha _n$ is an arbitrary real number, integer or noninteger,
positive or negative. Following the method of the algebraic self-similar
renormalization \cite{G1}, we define the algebraic transform 
\begin{equation}
\label{f2}P_k(x,s)\equiv x^s\ p_k(x)=\sum_{n=0}^ka_n\ x^{s+\alpha _n}, 
\end{equation}
where $s$ is real. Rather than constructing a trajectory in the functional
space of the initial approximations, the idea behind the introduction of the
transform (\ref{f2}) is to deform smoothly the initial functional space of
the approximations $p_k(x)$ in order to obtain a faster and better
controlled convergence in the space of the modified functions $P_k(x,s)$.
This convergence can then be mapped back to get the relevant estimations and
predictions. The exponent $s$ depends on $x$ in general and will be acted
upon as a control function in order to accelerate convergence.

Then, by means of the equation $P_0(x,s)=a_0\ x^{s+\alpha _0}=\varphi$, we
obtain the expansion function $x(\varphi ,s)=\left( \frac \varphi
{a_0}\right) ^{1/\left( s+\alpha _0\right)}$. Substituting the latter into (
\ref{f1}), we have 
\begin{equation}
\label{f5}y_k(\varphi ,s)\equiv P_k(x(\varphi ,s),s)=\sum_{n=0}^ka_n\ \left(
\frac \varphi {a_0}\right) ^{\left( s+\alpha _n\right) /\left(
s+\alpha_0\right) }. 
\end{equation}
The family $\{y_k\}$ of transforms (\ref{f1}) is called the approximation
cascade, since its trajectory $\{y_k(\varphi ,s)\mid k=0,1,2...\}$ is
bijective to the sequence $\{P_k(x,s)\mid k=0,1,2...\}$ of approximations (
\ref{f2}). A cascade is a dynamical system in discrete time $k=0,1,2...,$
whose trajectory points satisfy the semigroup property $y_{k+p}(\varphi,s)=
y_k(y_p(\varphi ,s),s).$ The physical meaning of the above semigroup
relation can be understood as the property of functional self-similarity
with respect to the varying approximation number. The self-similarity
relation is a necessary condition for the fastest convergence criterion.

For the approximation cascade $\{y_k\},$ defined by transform (\ref{f5}),
the cascade velocity is 
\begin{equation}
\label{f6}v_k(\varphi ,s)\equiv y_k(\varphi ,s)-y_{k-1}(\varphi ,s)=a_k\
\left( \frac \varphi {a_0}\right) ^{\left( s+\alpha _k\right) /\left(
s+\alpha _0\right) }. 
\end{equation}
This is to be substituted into the evolution integral 
\begin{equation}
\label{f7}\int_{P_{k-1}}^{P_k^{*}}\frac{d\varphi }{v_k(\varphi ,s)}=\tau , 
\end{equation}
in which $P_k=P_k(x,s)$ and $\tau $ is the minimal time needed for reaching
a fixed point $P_k^{*}=P_k^{*}(x,s,\tau ).$ Integral (\ref{f7}) with
velocity (\ref{f6}) yields 
\begin{equation}
\label{f8}P_k^{*}(x,s,\tau )=\left[ P_{k-1}^{-\nu }(x,s)-\frac{\nu \ a_k\tau 
}{a_0^{1+\nu }}\right] ^{-1/\nu }, 
\end{equation}
where $\nu =\nu _k(s)\equiv \frac{\alpha _k-\alpha _0}{s+\alpha _0}$. Taking
the algebraic transform inverse to (\ref{f2}), we find 
\begin{equation}
\label{f10}p_k^{*}(x,s,\tau )\equiv x^{-s}P_k^{*}(x,s,\tau )=\left[
p_{k-1}^{-\nu }(x)-\frac{\nu \ a_k\tau }{a_0^{1+\nu }}x^{s\nu }\right]
^{-1/\nu }. 
\end{equation}
Exponential renormalization \cite{G2,G4} corresponds to the limit $
s\rightarrow \infty ,$ at which $\lim _{s\rightarrow \infty }\nu _k(s)=0,\
\lim _{s\rightarrow \infty }s\nu _k(s)=\alpha _k-\alpha _0.$ Then (\ref{f10}
) gives 
\begin{equation}
\label{f11}\lim _{s\rightarrow \infty }p_k^{*}(x,s,\tau )=p_{k-1}(x)\exp
\left( \frac{a_k}{a_0}\tau x^{\alpha _k-\alpha _0}\right) . 
\end{equation}
Accomplishing exponential renormalization of all sums appearing in
expression of type (\ref{f11}), we follow the bootstrap procedure \cite{G2}
according to the scheme $p_k(x)\rightarrow p_k^{*}(x,s,\tau )\rightarrow
F_k(x,\tau _1,\tau _2,...,\tau _k)$, with $k\geq 1.$

Let us mention a recent innovation \cite{G6} that may improve significantly
the convergence of the method, based on the determination of the control
parameters from the knowledge of some moments of the function to
reconstruct. Let us assume that we can obtain the first $j-1$ moments $\mu
_i,i=1,2..,j$ of the sought function $\phi (t),$in some interval $T,$
\begin{equation}
\label{27}\mu _i=\int_0^Tt^{i-1}\phi (t)dt, 
\end{equation}
so that for j=2 both zero and first moments are available etc... One can
condition the control parameters $\tau _{1,}\tau _{2,..}\tau _j$ as follows 
\begin{equation}
\label{28}\int_0^Tf_j^{*}(t,\tau _{1,}\tau _{2,}...,\tau _{\
j-1})t^{i-1}dt=\mu _i, 
\end{equation}
Based on these conditions, one can attempt to solve two different problems,
the first one corresponds to an approximate reconstruction of the function $
\phi (t)$ within the same interval [0,T] where moments are given or
measured. The second problem consists in extrapolating to $t>T$. It is also
possible to use an hybrid approach, where some controls are obtained from
the agreement with expansion, while the remaining ones are found from the
conditions on moments.

\section{Resummation of lower order expansions in $\sigma _Y^2$.}

\subsection{What can be extracted from the expansion of $K_{ef}^{[1]}$}

Assume that the following extremely short expansion has been obtained, 
\be
K(\sigma_Y)\simeq 1-a\sigma _Y^2,~~~~~~~~ a=\frac 1d-\frac 12 ~~~~~~~~~
(\sigma _Y^2\rightarrow 0)~. 
\ee
Consider the case $\ a>0.$ In order to find the behavior of $K(\sigma _Y)$
for arbitrary $\sigma _Y^2$, we continue it from the region of $\sigma
_Y^2\rightarrow 0$ self-similarly, along the most stable trajectory, with
the crossover index $s$, determined by the condition of the minimum of the
multiplier \cite{G1,G2,G11} \be
m(\sigma _Y,s)=1-a\sigma _Y^2\frac{1+s}s, \ee
from where 
$$
s(\sigma _Y)=a\sigma _Y^2\left( 1-a\sigma _Y^2\right) ^{-1},\ \ \ \sigma
_Y<a^{-1/2}, 
$$
$$
s\rightarrow \infty ,\ \ \sigma _Y\geq a^{-1/2}, 
$$
corresponding to the self-similar approximation 
\be
K^{*}(\sigma _Y)=\left( \frac{s(\sigma _Y)}{s(\sigma _Y)+a\sigma _Y^2}
\right) ^{s(\sigma _Y)}=\left( 2-a\sigma _Y^2\right) ^{\frac{a\sigma _Y^2}{
a\sigma _Y^2-1}},\ \ \sigma _Y <a^{-1/2}~, \label{jgwl} 
\ee
\be
K^{*}(\sigma _Y)=\exp (-a\sigma _Y^2),\quad \sigma _Y\geq a^{-1/2}\ . 
\label{nnmask}
\ee
This suggests one self-similar expression (\ref{jgwl}) up to $\sigma
_{Y0}=a^{-1/2}$, and another (\ref{nnmask}), exponentially ``soft'', above
this value. Strictly speaking, formula (\ref{jgwl}) is applicable for $
\sigma _Y$ only up to $(2/a)^{1/2}$, where it predicts a spurious zero of
conductivity. In this particular case, the self-similar approximation
plausibly reconstructs the exponential function for arbitrary $\sigma _Y$,
even in the absence of any a priori assumption on the asymptotic behavior at 
$\sigma _Y\rightarrow \infty $.

For negative $a\ \left( d>2\right)$, there is no limiting value $\sigma
_{Y0} $ and this yields \be
K^{*}(\sigma _Y)=\left( \frac{s(\sigma _Y)}{s(\sigma _Y)+a\sigma _Y^2}
\right) ^{s(\sigma _Y)}~, \label{mgmwl} \ee
which can be used for arbitrary $\sigma _Y$. This expression appears to be
more stable than the previously proposed exponential solution $\exp
(-a\sigma _Y^2)$ for arbitrary $\sigma _Y$, as indicated by the analysis of
the multipliers. Expression (\ref{mgmwl}) also predict a smaller
conductivity than the exponential function for arbitrary $\sigma _Y$
suggesting that, for the most interesting case $d=3$, the ansatz equation (
\ref{k}) should be replaced. Analysis of higher-order expansions will
provide more details on the sought function.

\subsection{Resummation of the second-order expansion in $\,\sigma _Y^2.$
One-parameter formula}

Available from\cite{dagan93} in the next order in $\sigma _Y^2$, we have 
\be
K_{ef}^{[2]}\,\equiv \,\,K_{ef}^{[1]}+\,K_{ef}^{(2)}\,=\,K_g\,\left[
1+\left( \frac 12-\frac 1d\right) \,\sigma _Y^2+\frac 12\left( \frac
12-\frac 1d\right) ^2\sigma _Y^4\right]~. \ee
Below, for simplicity, we apply our resummation technique to the
dimensionless quantity \be
K(z)\simeq 1+a_1z+a_2z^2,\quad z\equiv \sigma _Y^2,\ \ a_1=\left( \frac
12-\frac 1d\right) ,\ a_2=\frac 12\left( \frac 12-\frac 1d\right) ^2~. 
\label{69} 
\ee

Application of accuracy-through-order conditions, or of the superexponential
approximants give, almost trivially, an exponential solution. Note that the
Pad\'e approximant available in this case, \be
P(z)=\frac{1+(-a_2/a_1\ +a_1)z}{1-a_2/a_1\ z}, \ee
for $d=3$, possesses a singularity at $z=12$, which is wrong.

The set of approximations to $K(z),\ $including the two starting terms from (
\ref{69}), can be written down as follows: 
$$
K_0=1, 
$$
$$
K_1=1+a_1z, 
$$
and the expression for the renormalized quantity $a_1^{*}$ can be readily
obtained: \be
K_1^{*}=\left( \frac{s_1}{s_1-a_1z}\right)  ^{s_1}\Longrightarrow \left( 
\frac{s_1}{-a_1}\right)  ^{s_1}\ z^{-s_1}\ (z\rightarrow \infty )~, \label
{70} \ee
where the stabilizer $s_1$ should be negative, if we want to reproduce in
the limit of $z\rightarrow \infty ,$ the correct, supposedly power-low
behavior of the conductivity. A different set of approximations, which does
not include the constant term from (\ref{69}) into the renormalization
procedure, has the form: \be
\overline{K_1}=a_1z, \ee
\be
\overline{K_2}=a_1z+a_2z^2, \ee
and applying the standard procedure of \cite{G2,G3}, we obtain 
\begin{equation}
\label{71}K_2^{*}=1+a_1z[1-\frac{a_2z}{a_1(1+s_2)}]^{-(1+s_2)}
\Longrightarrow (-\frac{a_2}{1+s_2})^{-(1+s_2)}\ a_1^{2+s_2}z^{-s_2}\
(z\rightarrow \infty )~.
\end{equation}
Demanding now that both (\ref{70}) and (\ref{71}) have the same power-law
behavior at $z\rightarrow \infty ,$ we find that 
$$
s_2=s_1\equiv s 
$$
Requiring now the fulfillment of the stability criteria for the two
available approximations in the form of the minimal-difference condition
(see section \ref{inrol}), we obtain the condition that the {\it negative}
stabilizer $s$ should be determined from the {\it minimum } of the
expression: \be
\left| \left[ \left( \frac{-a_2}{1+s}\right) ^{-(1+s)}\ a_1^{(2+s)}-\left(
\frac s{-a_1}\right) ^s\right] \right|  ~. \ee
Generally speaking, it is sufficient to ask for an extremum of this
difference.

In the case of $d=3$, the maximum is located at the point $s=-1.218$. The
final formulae have the following form: \be
K_1^{*}(z)=\left( \frac s{s-a_1z}\right) ^s, \ee
\be
K_2^{*}(z)=1+a_1z\left[ 1-\frac{a_2z}{a_1(1+s)}\right] ^{-(1+s)}\ . \ee
This last formula gives a lower bound for conductivity while the upper bound
is simply $exp(a_1z)$ (equation (\ref{k})) which, in this case, is the only
available ``factor''-approximant based on all available (three) terms from
the expansion. The corresponding multiplier is \be
M_2^{*}(z)=\frac{a_1(1+s)+a_2z\ s}{a_1(1+s)-a_2z}\left( 1-\frac{a_2}{a_1(1+s)
}z\right) ^{-(s+1)} \ee
and the weighted average \cite{G5,G6} is given by \be
C(z)=\frac{1+K_2^{*}(z)\left| M_2^{*}(z)\right| ^{-1}}{\exp (-a_1z)+\left|
M_2^{*}(z)\right| ^{-1}}, \ee
providing the one-parametric formula for 3d-conductivity. See Fig.1 for a
comparison of different formulas for the conductivity as a function of the
variance $z\equiv \sigma _Y^2$ defined in equation (\ref{69}). The solid
line corresponds to the average $C(z)$, while the dotted line presents $
K_2^{*}(z)$. The dashed line is the celebrated exponential
Landau-Lifshitz-Matheron (LLM) conjecture. The dash-dotted line corresponds
to the result of resummation based on first-order expansion, $K^{*}(\sigma
_Y)$.

\subsection{Resummation of the third-order expansion in $\sigma _Y^2.$
Two-parameters formula}

The expansion to the next order is given by \cite{dewit95}, 
\be
K(z)\simeq 1+a_1z+a_2z^2+a_3z^3,\quad \ a_3(Z)=\frac 16\left( \frac 12-\frac
1d\right) ^3-Z ~,
\label{mgjwkllw}
\ee
where $Z=0.0042/3$ (in the case of a Gaussian covariance ), or $Z=0.0014/3$
(in the case of an exponentially decaying covariance).

Using all terms from the expansion, we can create the following ``odd''
factor-approximant \cite{G8}, 
\be
K_3^{*}(z,Z)=1+a_1z\left( 1-\frac{a_2}{a_1(s_2(Z)+1)}z\right)
^{-(s_2(Z)+1)},\quad s_2(Z)=-2\frac{a_2^2-a_3(Z)a_1}{a_2^2-2a_3(Z)a_1}~, 
\label{mjgsl} 
\ee
while $K_2^{*}(z)=\exp (a_1z)$, which recovers expression (\ref{k}). The
approximant $K_2^{*}(z)$ gives an upper bound for the conductivity
coefficient, while $K_3^{*}(z,Z)$ given by (\ref{mjgsl}) provides a lower
bound. The corresponding multipliers can be readily written down, \be
M_3^{*}(z,Z)=\frac{a_1(1+s_2(Z))+a_2z\ s_2(Z)}{a_1(1+s_2(Z))-a_2z}\left( 1- 
\frac{a_2}{a_1(s_2(Z)+1)}z\right) ^{-(s_2(Z)+1)}, \ee
\be
M_2^{*}(z)=\exp (a_1z)~. \ee
This allows us to obtain the weighted average of the conductivity
coefficient \be
C(z,Z)=\frac{1+K_3^{*}(z,Z)\left| M_3^{*}(z,Z)\right| ^{-1}}{\exp
(-a_1z)+\left| M_3^{*}(z,Z)\right| ^{-1}}~, \ee
providing a two-parameters formula for the 3d-conductivity. The results for $
C(z,Z)$ for Gaussian and exponential covariances are shown in Fig. 2, with
the dotted line for the exponential case and with the solid line for the
Gaussian case. These results are compared with the Landau-Lifshitz-Matheron
(LLM) shown with dashed line.

Numerical data in the exponential case are available till $z=7$ \cite
{neumanorrpr} and in the Gaussian case up to $z=6$ \cite{Dykaarfged}. Our
results suggest that all formulas based on the expansion on $\sigma_Y^2-$ up
to the third order underestimate the conductivity and more terms are needed
to improve the accuracy of the resummed expressions.

A different approach aimed at increasing the accuracy consists in getting
expressions for the conductivity in the limit of $\sigma _Y^2\rightarrow
\infty $, for instance from expansions in inverse powers of $\sigma _Y^2$,
It is known (see Ref.\cite{G9,G10,G11}) that when the asymptotic form of the
solution is known, even only qualitatively, the formulas for the sought
function can be improved very significantly. Even the knowledge of the
leading power in the limit of large $\sigma _Y^2$ would be of utmost
importance.

\section{$1/d$-expansion and resummation}

\subsection{One-parametric case (d=3)}

Expression (\ref{69}) for $K(z)$ in the second order of perturbation theory
can be re-written in the form of an expansion in the parameter $1/d$ with
coefficients dependent on $z$, \be
K(m)\simeq b_0(z)+b_1(z)m+b_2(z)m^2,\quad m\equiv 1/d; \ee
\be
b_0(z)=1+\frac z2+\frac{z^2}8,\quad b_1(z)=-z-\frac{z^2}2,\quad b_2(z)=\frac{
z^2}2~. \ee
The theory of self-similar super-exponential approximants \cite{G2,G4,G6,G7}
then provides the following approximant \be
K_2^{*}(m)=b_0\exp \left( \frac{b_1}{b_0}\tau _1m\exp \left( \frac{b_2}{b_1}
\tau _2m\right) \right) ,\ \quad \tau _1=1,\ \tau _2=1-\frac{b_1^2}{2b_0b_2}
~. \ee
$K_2^{*}(m)$ is located within the bounds given by $K_2^{*}(z)$ and exp($
a_1z $) and provides a one-parametric formula for the $d=3$-conductivity, as
shown in Fig.~3. $K_2^{*}(m)$ (dotted line) appears to be located within the
bounds outlined by $C(z)$ (dashed) and $exp$($a_1z$) (solid line) and
provides a one-parametric formula for the conductivity in three dimensions.
The perturbative expression $K(m)$ (dashed-dot line) is shown as well for
comparison.

\subsection{Two-parametric case (d=3)}

Expression for $K(z)$ (\ref{mgjwkllw}) in the third order of perturbation theory
can also be re-written in the form of an $1/d$-expansion with coefficients
dependent on $z$ and $Z$,\be
K(m,Z)\simeq b_0(z,Z)+b_1(z)m+b_2(z)m^2+b_3(z)m^3; \ee
\be
b_0(z,Z)=1+\frac z2+\frac{z^2}8+(\frac 1{48}-Z)z^3,\quad b_1(z)=-z-\frac{z^2}
2-\frac{z^3}8,\quad b_2(z)=\frac{z^2}2+\frac{z^3}4, \ee
\be
b_2(z)=\frac{z^2}2+\frac{z^3}4,\qquad b_3(z)=-\frac{z^3}6. \ee
We apply the technique of self-similar superexponential function in its
variant detailed in \cite{G6,G7} , giving the following approximants \be
K_2^{*}(m,Z,\tau _1,\tau _2)=b_0\exp \left( \frac{b_1}{b_0}\tau _1m\exp
\left( \frac{b_2}{b_1}\tau _2m\right) \right) , \ee
\be
K_3^{*}(m,Z,\tau _1,\tau _2,\tau _3)=b_0\exp \left( \frac{b_1}{b_0}\tau
_1m\exp \left( \frac{b_2}{b_1}\tau _2m\exp \left( \frac{b_3}{b_2}\tau
_3m\right) \right) \right) , \ee
\be
\tau _3=\frac 16\frac{\left( -b_1^4+6b_0^2b_1b_3-6\tau _2b_0b_1^2b_2-3\tau
_2^2b_0^2b_2^2\right) }{\tau _2b_0^2b_1b_3}~. \ee
In order to check whether the sequence\ of $K_j^{*}$\ converges, we study
their mapping multipliers, $M_j^{*}(t,\tau _1,\tau _2,\ldots \tau _j)$
defined as \be
M_j^{*}(m,Z,\tau _1,\ldots ,\tau _j)\equiv \frac 1{b_1}\ \frac \partial
{\partial t}\ K_j^{*}(m,Z,\tau _1,\ldots ,\tau _j)\ . \ee
This definition of the multipliers allows us to compare the convergence of
the expansion and of the renormalized expressions, making clear what can be
expected a priori.

This provides a matrix of self-similar approximants, indexed by the order $j$
and by the number of control parameters, \be
K_{21}^{*}(m,Z)=K_j^{*}(m,Z,\tau _1,1),\qquad
K_{22}^{*}(m,Z)=K_j^{*}(m,Z,\tau _1,\tau _2),\qquad \ \quad \qquad \qquad
\quad \qquad \qquad \quad \ee
\be
K_{31}^{*}(m,Z)=K_3^{*}(m,Z,\tau _1,1,1),\quad
K_{32}^{*}(m,Z)=K_3^{*}(m,Z,\tau _1,\tau _2,1),\quad
K_{33}^{*}(m,Z)=K_3^{*}(m,Z,\tau _1,\tau _2,\tau _3), \ee
Approximants $K_{22}^{*}(m,Z)$ and $K_{32}^{*}(m,Z)$ form the closest pair.
Their average \be
K^{*}(m,Z)=\frac{K_{22}^{*}(m,Z)\left| M_{22}^{*}(m,Z)\right|
^{-1}+K_{32}^{*}(m,Z)\left| M_{32}^{*}(m,Z)\right| ^{-1}}{\left|
M_{22}^{*}(m,Z)\right| ^{-1}+\left| M_{32}^{*}(m,Z)\right| ^{-1}} ~. \ee
is located within the bounds given by $K_3^{*}(z,Z)$ (\ref{mjgsl}) and exp($
a_1z$) and provides a useful formula for the conductivity coefficient. The
results for the exponential and Gaussian covariances are shown in Figures 4
and 5 respectively. In the exponential case, good agreement between $
K^{*}(m,Z)$ (dash-dot) and the Landau-Lifshitz-Matheron (LLM) conjecture
(solid) remains valid till $z\approx 11$. The average behavior appears to be
located within the bounds outlined by $K_{22}^{*}(m,Z)$ (dashed) and $
K_{32}^{*}(m,Z)$ (dotted) and provides a reasonable formula for the
conductivity. Numerical data are available till $z=7$ \cite{neumanorrpr} and
they agree well with our lower bound.

In the Gaussian case $K_{32}^{*}(m,Z)$ (dotted line) gives an approximation
which is closer to the Landau-Lifshitz-Matheron (LLM) conjecture (solid)
than $K_{22}^{*}(m,Z)$ (dashed line). In this case, numerical results are
available up to $z=6$ \cite{Dykaarfged}.

We conclude, tentatively, that all formulas based on $1/d-$expansion in the
third order provide rather accurate expressions for conductivity for small
and moderate variances and disagree with the LLM-conjecture for very large
variances. To the best of our knowledge, the region of large variances is
not accessible by other techniques, numerically or theoretically. All
formulas based on the novel proposed $1/d-$expansion in the third order
provide rather accurate expressions for the conductivity coefficient. We
note that $1/d$-expansions may be faster converging than the original
expansion in variances and should be investigated future.

\section{Future directions}

We have shown that it is possible to extend the moment equation approach
using the self-similar functional renormalization method
in order to provide a stable and robust estimation of the transport
variables of the three-dimensional medium at large values of the
perturbation parameter $\sigma _Y^2$. 

Future directions include the
extension of the self-similar functional renormalization method to go beyond
the characterization of the heterogeneity solely in terms of the variance
and consider also the dependence of the transport properties with respect to
the skewness (third normalized cumulant) and kurtosis (fourth normalized
cumulant).

{\bf Acknowledgment}: We are grateful to D.M.~Tartakovsky for introducing us
into the subject and for useful discussions.

\pagebreak

{\bf Figures Captions}

\vskip 1cm

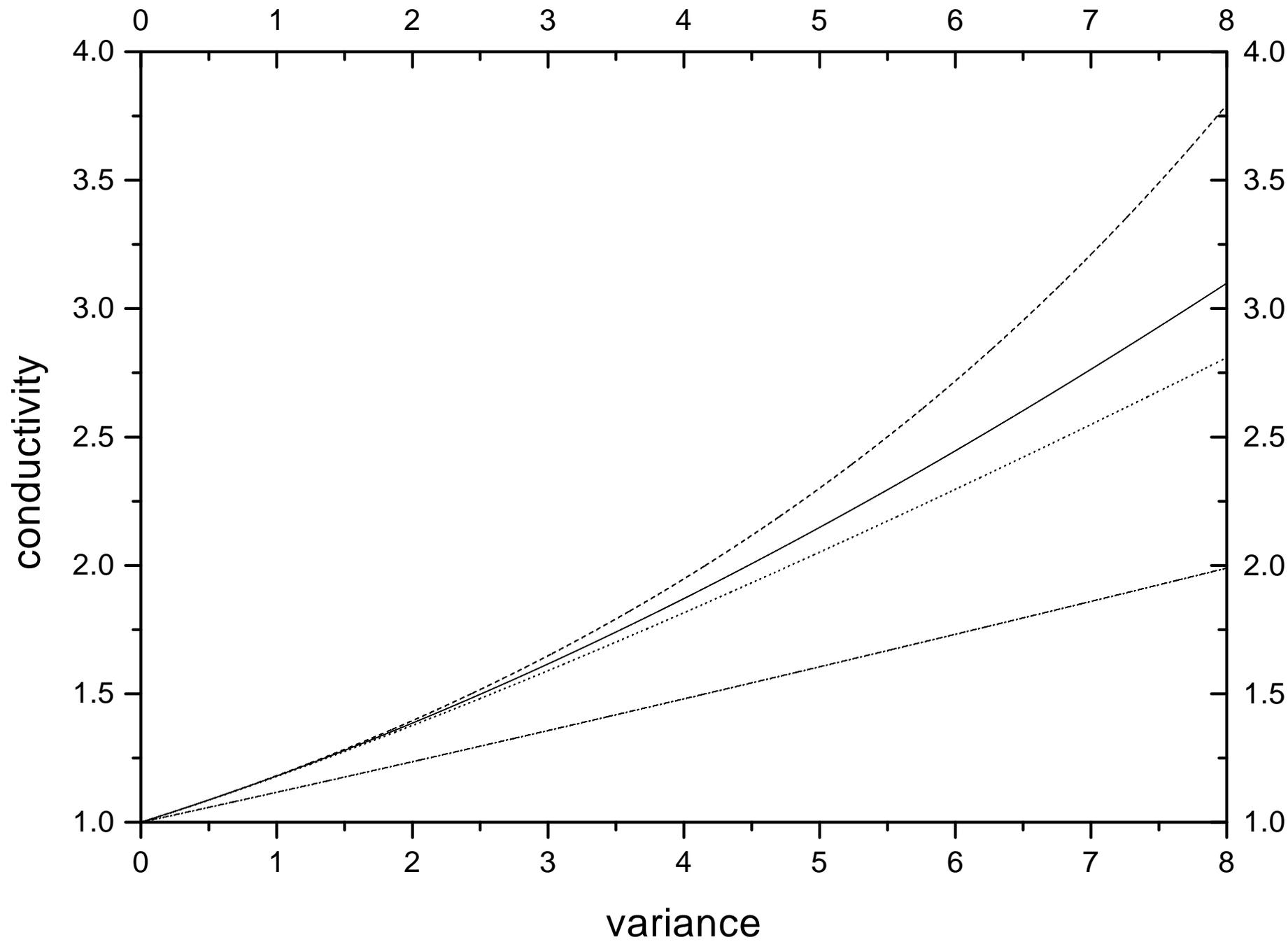
Figure 1. Dependence of different estimators of the conductivity as a
function of the variance $z\equiv \sigma _Y^2$ defined in equation (\ref{69}%
). The solid line corresponds to $C(z)$, the dotted line shows $K_2^{*}(z)$.
The dashed line is the Landau-Lifshitz-Matheron (LLM) conjecture. $%
K^{*}(\sigma _Y)$ (dash-dotted line) corresponds to the result of
resummation based on the first-order expansion.

\vskip 0.5cm Figure 2. Dependence of the conductivity $C(z,Z)$ as a function
of the variance $z\equiv \sigma _Y^2$ defined in equation (\ref{69}) for
Gaussian and exponential covariances, shown with solid and dotted lines
respectively. For comparison, the Landau-Lifshitz-Matheron (LLM) conjecture
is shown in dashed lines.

\vskip 0.5cm Figure 3. As a function of the variance $z\equiv \sigma _Y^2$
defined in equation (\ref{69}), this figure shows a comparison between the
conductivity $K_2^{*}(m)\ $(dotted line) with $C(z)$ (dashed) and with exp($%
a_1z$) (solid line). The Perturbative expansion $K(m)$ is shown with the
dashed-dot line.

\vskip 0.5cm Figure 4. Exponential covariance: $K^{*}(m,Z)$ (dash-dot) is
compared with the Landau-Lifshitz-Matheron (LLM) conjecture (solid line).
Two approximants $K_{22}^{*}(m,Z)$ (dashed) and $K_{32}^{*}(m,Z)$ (dotted)
are shown as well.

\vskip 0.5cm Figure 5. Gaussian case: Approximant $K_{32}^{*}(m,Z)$ (dotted)
compared with the Landau-Lifshitz-Matheron (LLM) conjecture (solid line).
The approximant $K_{22}^{*}(m,Z)$ (dashed line) is shown as well.

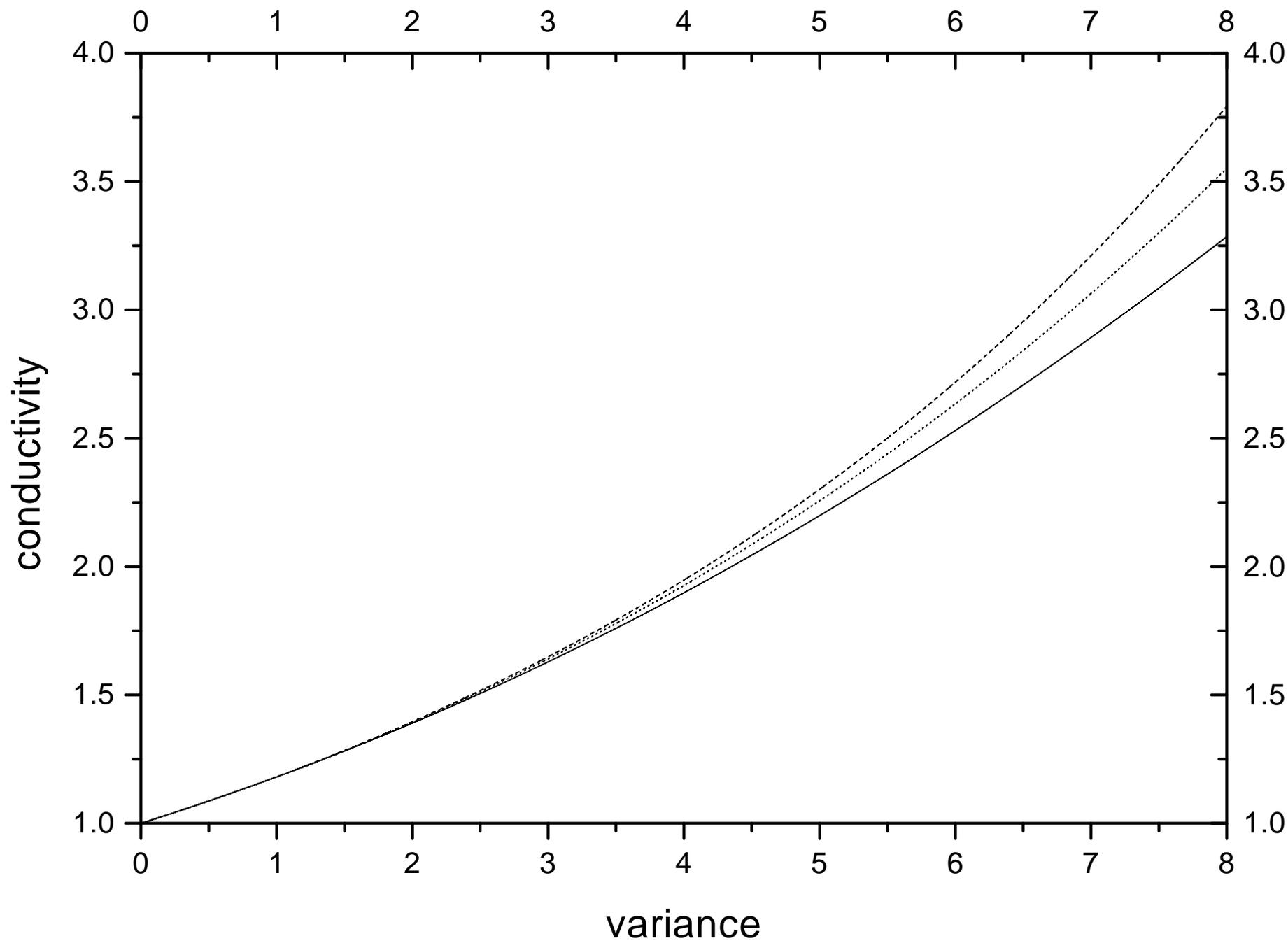
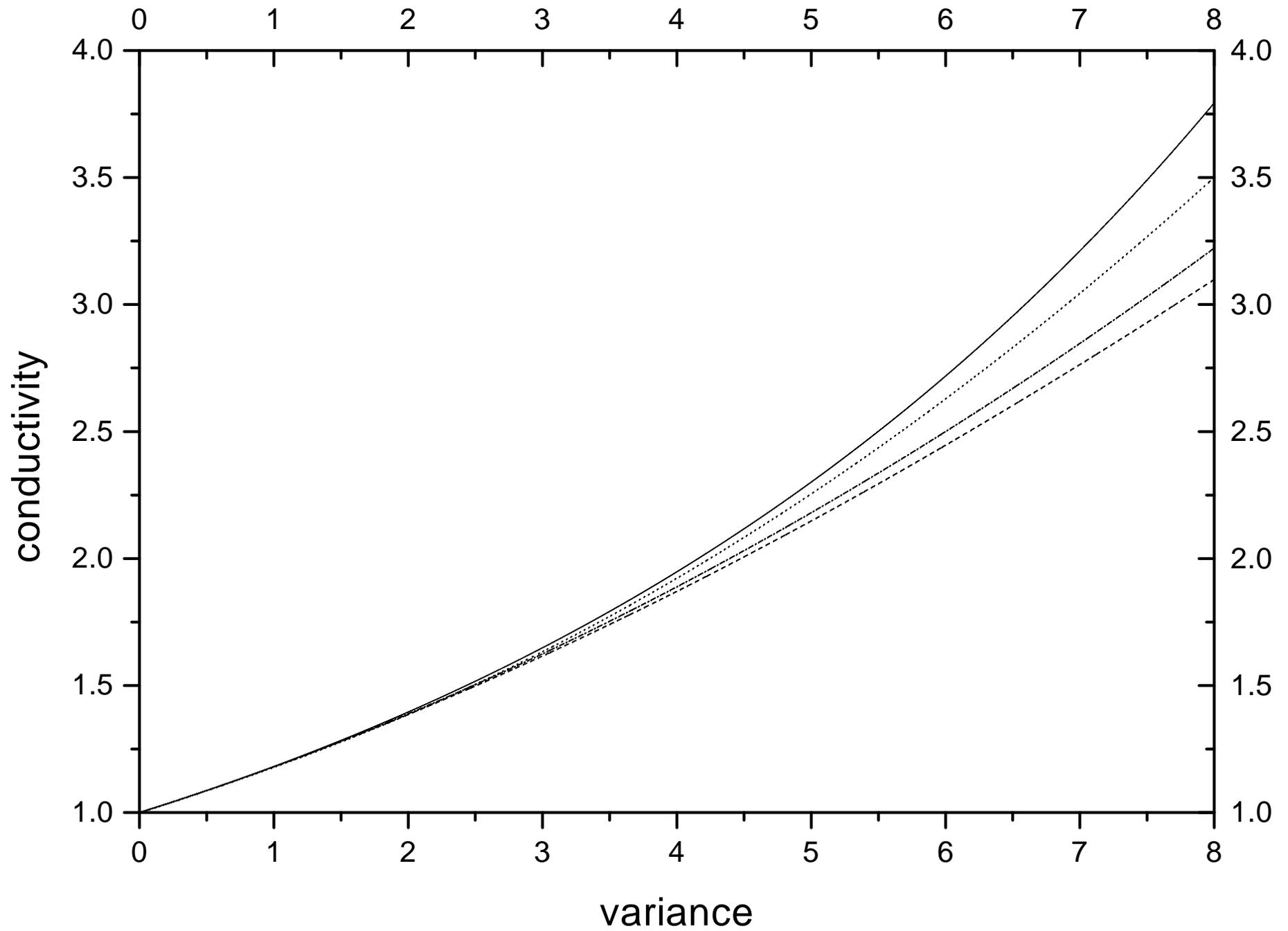
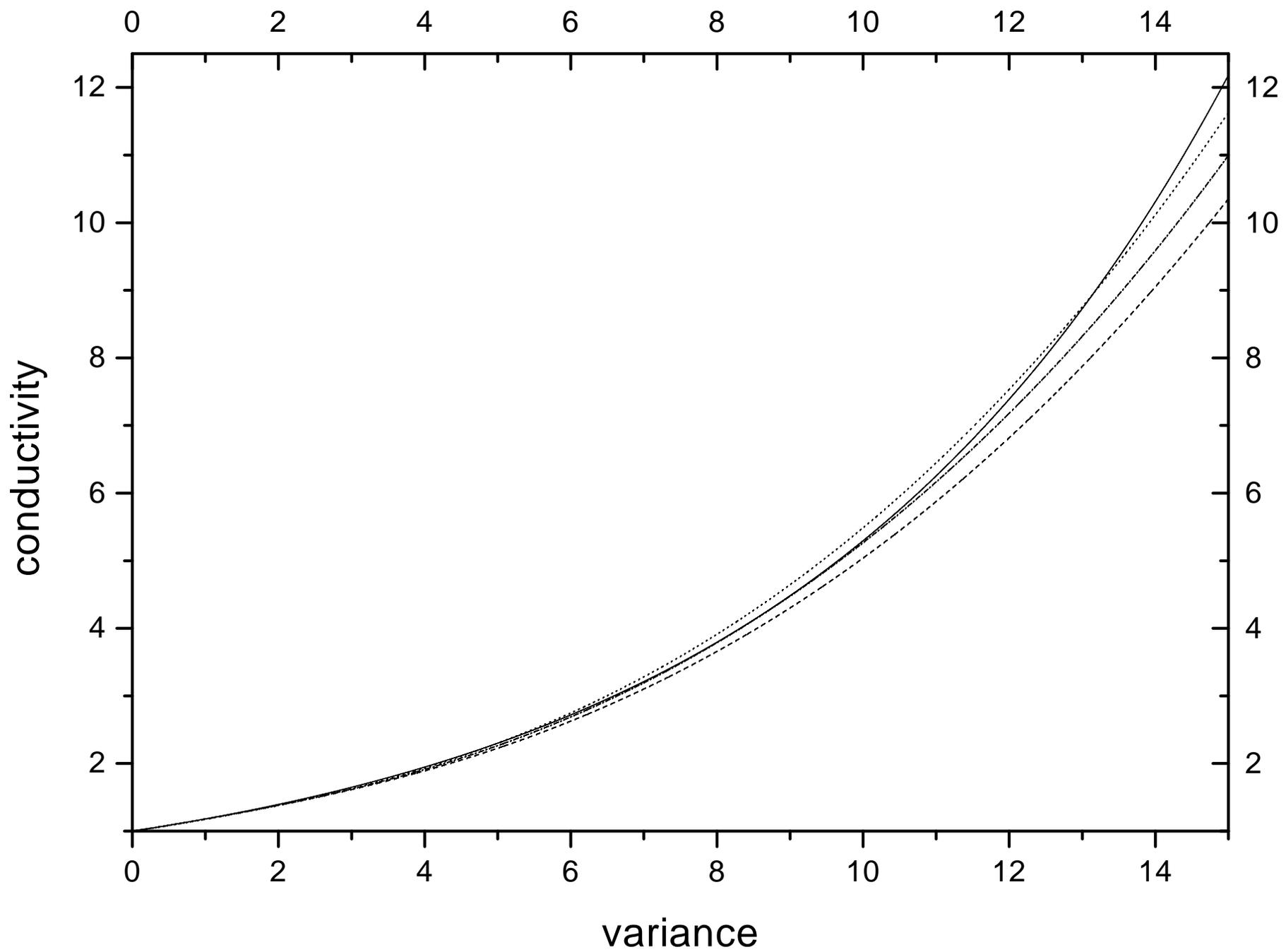
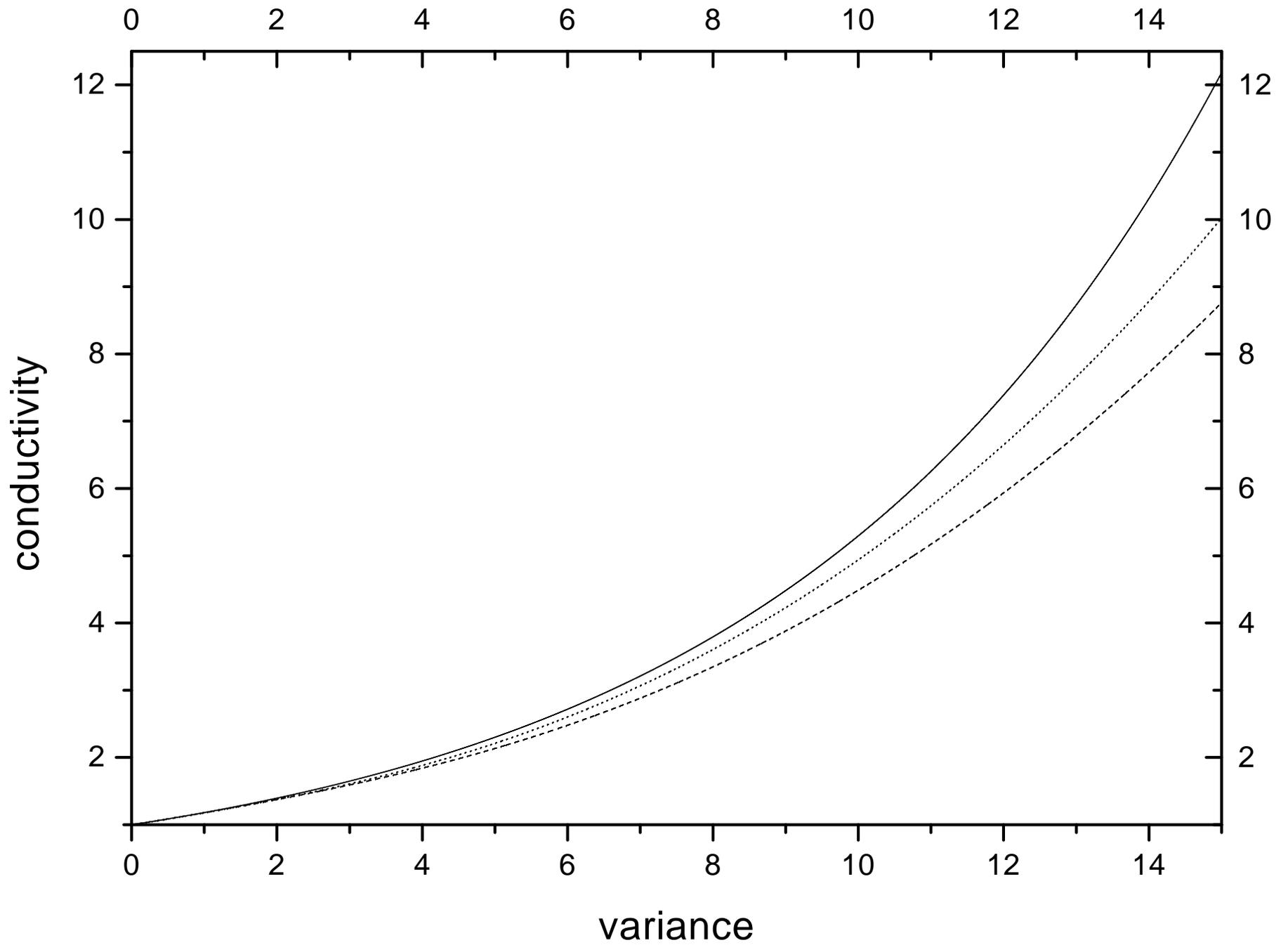
\end{document}